\documentclass[aps,twocolumn,pra,superscriptaddress,amsmath,showpacs,tightenlines]{revtex4-1}
\usepackage{amssymb}
\usepackage{braket}
\usepackage{amsmath}
\usepackage{graphicx}
\usepackage{subfigure}
\usepackage{natbib}
\usepackage{epsfig}
\usepackage{amsfonts}
\usepackage{mathrsfs}
\usepackage{CJK}
\usepackage{xcolor}
\usepackage{bm}
\usepackage[toc,page,title,titletoc,header]{appendix}
\begin{document}
\title{Enantiodetection in a cavity QED setup with finite chiral molecules}
\author{Xiang Guo}
\affiliation{Center for Quantum Sciences and School of Physics, Northeast Normal University,
Changchun 130024, China}
\author{Xiaojun Zhang}
\affiliation{Center for Quantum Sciences and School of Physics, Northeast Normal University,
Changchun 130024, China}
\author{Yong Li}
\email{yongli@hainanu.edu.cn}
\affiliation{Center for Theoretical Physics \& School of Physics and Optoelectronic Engineering, Hainan University, Haikou 570228, China}
\author{Zhihai Wang}
\email{wangzh761@nenu.edu.cn}
\affiliation{Center for Quantum Sciences and School of Physics, Northeast Normal University,
Changchun 130024, China}
\begin{abstract}
We investigate enantiodetection for both a single cyclic three-level chiral molecule and finite ensembles of such molecules by monitoring the steady-state intracavity photon number in a cavity-QED platform. Our scheme exploits the intrinsic global $\pi$-phase difference between opposite enantiomers to engineer destructive and/or constructive interference pathways, enabling a direct readout of enantiomeric excess with an error below $5\%$. To capture mesoscopic many-molecule effects beyond mean field while avoiding brute-force master-equation simulations, we employ a generalized discrete truncated Wigner approximation, which is well suited for systems with many yet finite molecules. These results pave the way for implementing enantiodetection in realistic quantum-optical settings.
\end{abstract}
\maketitle
\section{Introduction}
Chiral molecules cannot be superimposed on their enantiomers by any combination of translations and rotations~\cite{RW1976}. As a result, left- and right-handed enantiomers are nearly indistinguishable in most physical properties yet can exhibit pronounced differences in physiological activity~\cite{NP1991,KB1997}. Consequently, enantiodetection~\cite{WZ2011,EH2012,AY2016,CY2019,YY2021,MR2022,YY2022,FZ2022} and enantiomeric conversion~\cite{MS2000,DG2001,CY2020,CY2021,CY2021_2} are central issues in pharmaceutical science. A wide variety of discrimination methods based on conventional spectroscopic techniques have been explored~\cite{PJ1985,RK1998,RB2005,SB2018}; however, they typically rely on magnetic-dipole and electric-quadrupole transitions between molecular energy levels, whose matrix elements are small, leading to intrinsically weak signals.

Recent progress in quantum light-matter interaction has enabled precise control of chemical reactions~\cite{RS1992,MY1996,DR1996,BJ2006,DC2008,LL2015,IR2024,JR2025}, motivating the use of quantized electromagnetic field as a new resource for enantiodetection~\cite{WZ2011,EH2012,AY2016,CY2019,YY2021,MR2022,YY2022,FZ2022}. In particular, for cyclic three-level ($\Delta$-type) chiral molecules in the microwave domain~\cite{YX2005,CY2018}, the product of the three electric-dipole coupling matrix elements carries a global $\pi$-phase difference between opposite enantiomers~\cite{PK2001,PK2003}. This phase property naturally enables interference-based detection schemes. Moreover, by driving electric-dipole transitions between the levels, the achievable signal strengths are substantially enhanced compared to approaches that rely on magnetic-dipole or electric-quadrupole transitions.

In this work, we consider chiral molecules coupled simultaneously to two classical fields and a quantized cavity mode in a cavity-QED setup where the cavity is furthermore coherently driven. The resulting chirality-dependent constructive and/or destructive interference between pathways, enabling enantiodetection of molecules via a simple readout of the steady-state intracavity photon number.

Previous quantum protocols for measuring enantiomeric excess typically assume the thermodynamic limit and a low-excitation regime~\cite{YY2021,YY2022}. For mixtures with a finite number of molecules-i.e., outside the thermodynamic limit-intermolecular correlations and fluctuations crucially affect the dynamics, invalidating mean-field treatments used in Refs.~\cite{YY2021,YY2022}. While quantum master equation simulations yield exact dynamics for small systems, the Hilbert space dimension grows exponentially with the number of molecules, which quickly becomes prohibitive. We therefore employ the generalized discrete truncated Wigner approximation (GDTWA)~\cite{SL2019,LP2021}, a semiclassical phase space method based on stochastic trajectory sampling that reduces the computational cost from exponential to approximately linear scaling with system size while retaining fluctuation effects beyond mean field. In practice, it samples the initial state from a `Wigner' distribution~\cite{AP2010,JS2015,JH2021,JH2022} and propagate an ensemble of trajectories, thereby capturing correlation-induced corrections beyond mean field.

Moreover, for finite-size mixtures we find that the amplitude of the molecule-mediated pathways depend sensitively on the number of molecules, rendering the intracavity photon number a direct proxy for the enantiomeric excess. In our simulations, the enantiomeric excess can be retrieved with a detection error below $5\%$. The operating range with maximal accuracy can be shifted on demand by tuning the drive strength.
\section{Model}
\begin{figure}
 \includegraphics[width=1\columnwidth]{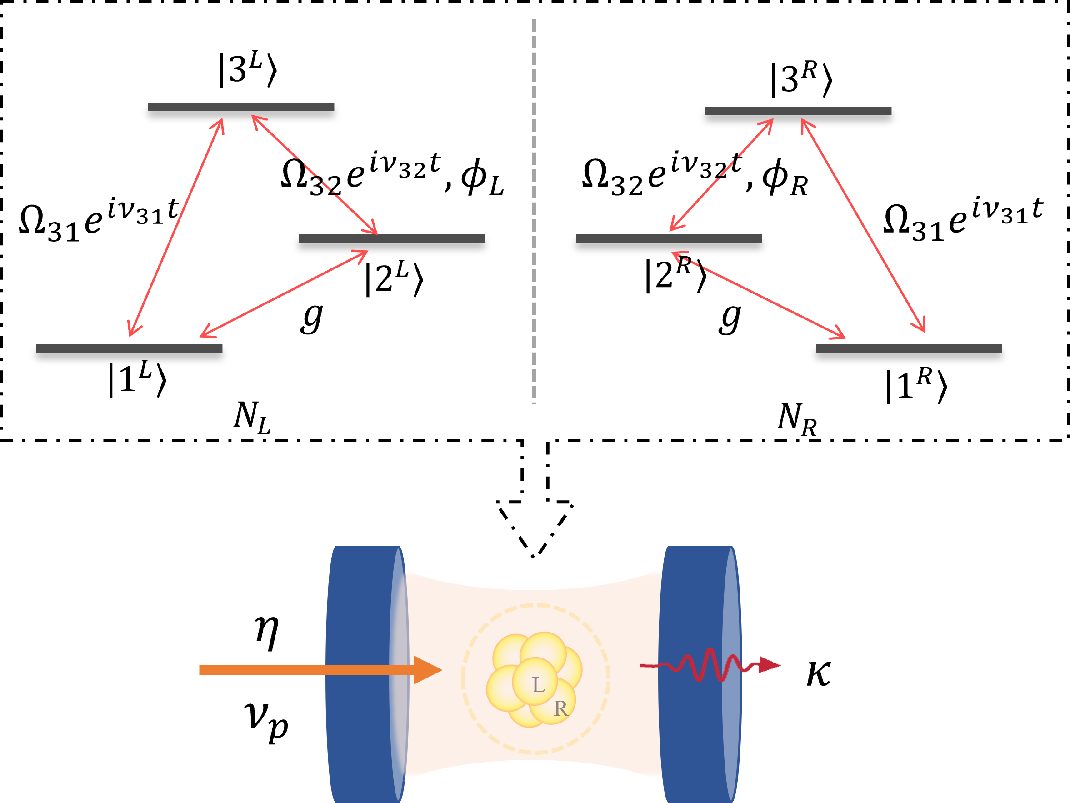}
 \caption{Schematic of the enantiodetection setup. A Fabry-P\'{e}rot cavity hosts $N_R$ right-handed and $N_L$ left-handed cyclic three-level ($\Delta$-type) molecules. Each molecule couples to two classical control fields with Rabi amplitudes $\Omega_{31}$ and $\Omega_{32}$ and to the quantized cavity mode with strength $g$. The closed-loop product of dipole couplings acquires a $\pi$-phase difference between opposite enantiomers. The cavity is driven coherently at frequency $\nu_{p}$ with amplitude $\eta$ and experiences photon loss at rate $\kappa$.}
 \label{Model}
\end{figure}
We consider a cavity-molecule model in which $N_L$ left-handed and $N_R$ right-handed chiral molecules couple to a single mode Fabry--P\'{e}rot cavity, as sketched in Fig.~\ref{Model}. Each enantiomer has three internal levels $\ket{i^{Q}}$ ($i=1,2,3$), where $Q\in\{L,R\}$ labels the chirality. Each molecule interacts with a quantized cavity mode and two classical control fields, forming a cyclic three-level ($\Delta$-type) configuration. The cavity is coherently driven with amplitude $\eta$ at frequency $\nu_p$. Within the rotating-wave approximation, the Hamiltonian is (set $\hbar=1$ throughout)
\begin{align}
\hat{H}'(t)=&\omega_{c}\hat{a}^{\dagger}\hat{a}+\Big\{\sum_{Q=L,R}\sum_{n_{Q}=1}^{N_{Q}}\big[\sum_{i=1,2,3}\omega_{i}\hat{\sigma}_{i,i}^{(n_{Q})}+\hat{H}^{(n_{Q})}\big]\nonumber\\
&+\eta(\hat{a}^{\dagger}e^{-i\nu_{p}t}+\hat{a}e^{i\nu_{p}t})\Big\}.
\end{align}
where $\hat{a}$ ($\hat{a}^\dagger$) annihilates (creates) a cavity photon of frequency $\omega_c$, and $\hat{\sigma}_{ij}^{(n_Q)}=\ket{i^{Q}_{n_Q}}\!\bra{j^{Q}_{n_Q}}$ acts on $n_Q$th molecule of chirality $Q$. The molecular bare energy of $\ket{i^{Q}_{n_Q}}$ is $\omega_i$, assumed identical for both enantiomers. The single-molecule interaction Hamiltonian reads
\begin{align}
&&\hat{H}^{(n_{Q})}=g\,\hat{\sigma}_{21}^{(n_{Q})}\hat{a}+\Omega_{32}\,e^{-i\nu_{32}t}\,e^{i\phi_{Q}}\,\hat{\sigma}_{32}^{(n_{Q})}\nonumber\\
&&+\Omega_{31}\,e^{-i\nu_{31}t}\,\hat{\sigma}_{31}^{(n_{Q})}+\text{H.c.},
\end{align}
where the transition $\ket{2^{Q}_{n_Q}}\!\leftrightarrow\!\ket{1^{Q}_{n_Q}}$ couples to the cavity mode with strength $g$, while $\ket{3^{Q}_{n_Q}}\!\leftrightarrow\!\ket{2^{Q}_{n_Q}}$ and $\ket{3^{Q}_{n_Q}}\!\leftrightarrow\!\ket{1^{Q}_{n_Q}}$ transitions are driven by classical fields at frequencies $\nu_{32}$ and $\nu_{31}$ with Rabi amplitudes $\Omega_{32}$ and $\Omega_{31}$, respectively. The overall loop phase $\phi_Q$ encodes the chirality, with $\Delta\phi=\phi_{R}-\phi_{L}=\pi$, which constitutes the essential distinction between the two enantiomers.

Under the three-photon resonance condition $\nu_{31}=\nu_{p}+\nu_{32}$, and moving to the interaction picture defined by
$\nu_{p}\hat{a}^{\dagger}\hat{a}
+\sum_{Q=L,R}\sum_{n_{Q}=1}^{N_{Q}}
\!\big[(\omega_{1}+\nu_{31})\hat{\sigma}^{(n_{Q})}_{33}
+\omega_{1}\hat{\sigma}^{(n_{Q})}_{11}
+(\omega_{1}+\nu_{31}-\nu_{32})\hat{\sigma}^{(n_{Q})}_{22}\big]$,
the Hamiltonian becomes time-independent and reads
\begin{align}
\hat{H} =& \Delta_{c}\hat{a}^{\dagger}\hat{a} + \sum_{Q=L,R}\sum_{n_{Q}=1}^{N_{Q}} \bigg[\Delta_{31}\hat{\sigma}_{3,3}^{(n_{Q})} + (\Delta_{31} - \Delta_{32})\hat{\sigma}_{2,2}^{(n_{Q})}\nonumber\\
&+\left(\Omega_{31}\hat{\sigma}_{3,1}^{(n_{Q})} + \Omega_{32}\hat{\sigma}_{3,2}^{(n_{Q})}e^{i\phi_{Q}} + g\hat{\sigma}_{2,1}^{(n_{Q})}\hat{a} + \text{H.c.}\right)\bigg]\nonumber\\
&+\eta(\hat{a}^{\dagger} + \hat{a}).
\label{H}
\end{align}
with detunings
$\Delta_{c}=\omega_{c}-\nu_{p}$,
$\Delta_{31}=\omega_{3}-\omega_{1}-\nu_{31}$,
and
$\Delta_{32}=\omega_{3}-\omega_{2}-\nu_{32}$.
Including cavity loss at rate $\kappa$, the system dynamics is governed by the master equation
\begin{align}
\frac{d\hat{\rho}}{dt}
= -i[\hat{H},\hat{\rho}]
+ \frac{\kappa}{2}\Big(2\hat{a}\hat{\rho}\hat{a}^{\dagger}
- \hat{a}^{\dagger}\hat{a}\hat{\rho}
- \hat{\rho}\hat{a}^{\dagger}\hat{a}\Big).
\label{ME}
\end{align}
\section{Enantiodetection}
\subsection{Single molecule}
\begin{figure}
 \includegraphics[width=1\columnwidth]{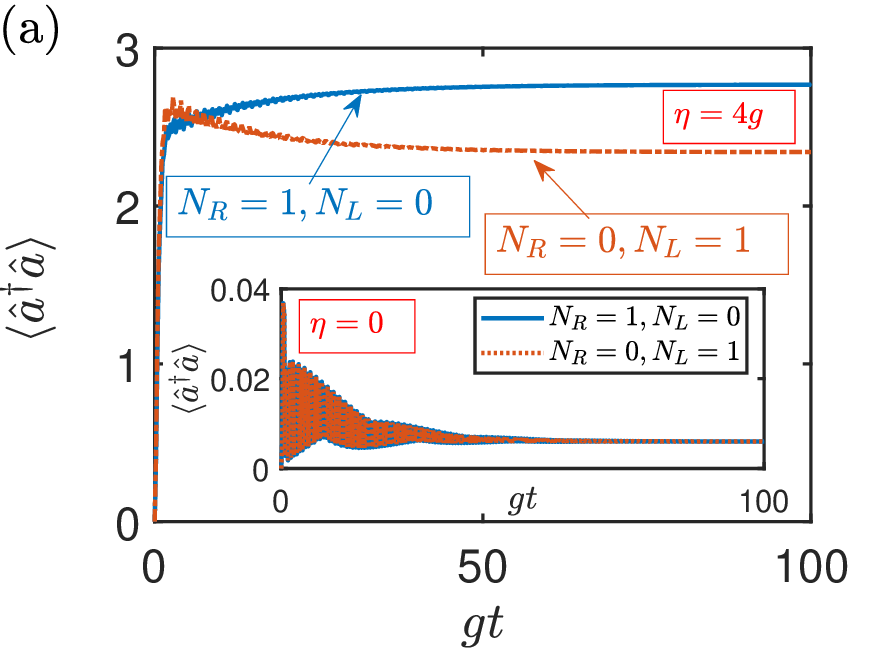}\\
 \includegraphics[width=1\columnwidth]{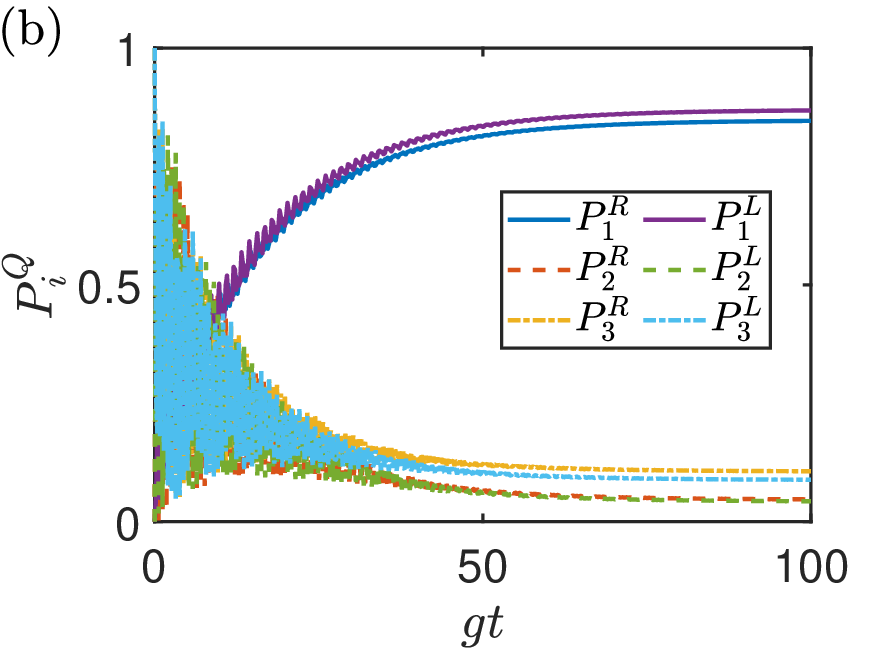}
 \caption{(a) and (b) photon and molecular dynamics, respectively. The parameters are set as $\Delta_{31}=\Delta_{32}=\Delta_{c}=0,\Omega_{32}=5g,\Omega_{31}=g,\kappa=5g,\eta=4g,\phi_{L}=0$, and $\phi_{R}=\pi$ in the main panel of (a) and (b). In the inset of (a), all parameters remain unchanged except for $\eta = 0$.}
 \label{discrimination}
\end{figure}

In this section, we analyze enantiodetection at the single-molecule level. In this case, the master equation in Eq.~(\ref{ME}) can be solved numerically. Fig.~\ref{discrimination}(a) shows the photon-number dynamics for a single molecule. When $\eta\neq 0$, the chirality manifests as a clear difference in the steady-state photon number. In contrast, in the absence of driving ($\eta=0$), the photon evolution in the two cases is indistinguishable.

We next attribute the difference in the steady-state photon number to interference between transition pathways. Before doing so, we plot the level populations $P_{i}^{Q}=\mathrm{Tr}\!\left[\ket{i^{Q}}\bra{i^{Q}}\hat{\rho}(t)\right]$~(Throughout this section we abbreviate $\ket{i^{Q}_{1}}$ as $\ket{i^{Q}}$) in Fig.~\ref{discrimination}(b) under finite driving. In both cases (single left- or right-handed molecule), the steady-state population of $\ket{3^{Q}}$ tracks the intracavity photon number: a larger average photon number correlates with a higher population in $\ket{3^{Q}}$. We therefore focus on the two interfering pathways that populate $\ket{3^{Q},n}$, where $\ket{n}$ denotes the cavity Fock state:
\begin{align}
1.&\quad \ket{3^{Q},n-1}\xrightarrow{\ \eta\ } -i\,\ket{3^{Q},n},\nonumber\\
2.&\quad \ket{3^{Q},n-1}\xrightarrow{\ \Omega_{32}\ } -i e^{-i\phi_{Q}}\ket{2^{Q},n-1}\nonumber\\
&\xrightarrow{\ g\ } - e^{-i\phi_{Q}}\ket{1^{Q},n}\xrightarrow{\ \Omega_{31}\ } i e^{-i\phi_{Q}}\ket{3^{Q},n}.
\label{single}
\end{align}

The first pathway is provided directly by the coherent cavity drive, while the second is mediated by the molecule. Each coherent coupling induces Rabi oscillations between the two connected states. For example, if the system is initialized in $\ket{3^{Q},n-1}$, the drive generates
\begin{equation}
\ket{\psi^{Q}(t)}=\cos(\eta t)\,\ket{3^{Q},n-1}-i\sin(\eta t)\,\ket{3^{Q},n},
\end{equation}
so that each such interaction contributes a phase factor of $-i$. For $\eta=0$, only the molecular-mediated pathway remains; no interference occurs, and the steady-state photon numbers for left- and right-handed cases are indistinguishable. By contrast, when $\eta\neq 0$, both the drive-induced and molecule-mediated pathways are present and interfere in a chirality-dependent manner. Specifically, for a single left-handed (right-handed) molecule, the two pathways in Eq.~(\ref{single}) interfere destructively (constructively), yielding a lower (higher) steady-state intracavity photon number. Thus, the coherent cavity drive enables single-molecule enantiodetection via steady-state photon-number readout, consistent with Fig.~\ref{discrimination}(a).
\subsection{Finite molecules}
Having established single-molecule enantiodetection, we now turn to mixtures comprising a finite number of molecules with unknown populations of left- and right-handed species. Specifically, we consider samples with $N_L$ left-handed and $N_R$ right-handed molecules and characterize the imbalance by the enantiomeric excess
\begin{equation}
\mathcal{P}=\frac{N_R-N_L}{N_R+N_L}.
 \end{equation}
We show below that, in this finite-size regime, the steady-state intracavity photon number provides a sensitive estimator of $\mathcal{P}$.

\subsubsection{Numerical approach: generalized discrete truncated Wigner approximation}
In the finite-molecule regime of interest, the Hilbert-space dimension scales as $3^{N_L+N_R}\!\times N_c$, where $N_c$ is the photon-number cutoff, rendering brute-force master equation simulations impractical once $N_L+N_R$ becomes large. At the same time, intermolecular correlations and quantum fluctuations are non-negligible, so mean-field descriptions fail to capture the dynamics of finite ensembles.

To access the finite-size regime beyond the thermodynamic limit--and outside the tractable range of brute-force master equation numerics--we employ the generalized discrete truncated Wigner approximation (GDTWA)~\cite{SL2019} (see Appendix~A for details).

(i) Starting from mean-field equations for local operators, GDTWA yields $\big[(D^{2}-1)(N_{L}+N_{R})+1\big]$ coupled differential equations, where $D$ is the number of internal levels per molecule ($D=3$ in this work). Thus, the computational cost changes from exponential in $N_{L}+N_{R}$ to approximately linear, greatly reducing complexity.

(ii) Quantum fluctuations are incorporated by sampling the initial state from a `Wigner' distribution for the operator averages, generating an ensemble of stochastic trajectories. Each trajectory evolves under the classical equations of motion, and adding a Gaussian stochastic term calibrated to the cavity loss rate to mimic dissipation-induced noise. This procedure captures fluctuation and correlation effects beyond mean field, albeit only to low order, so GDTWA is most reliable when quantum fluctuations are moderate.

(iii) Symmetrically ordered expectation values are obtained by averaging the corresponding classical variables over the trajectory ensemble. For arbitrary operators $\hat{A}$ and $\hat{B}$,
\begin{align}
\langle \{\hat{A}\hat{B}\}_{s}\rangle(t)
= \frac{1}{n_{t}}\sum_{i=1}^{n_{t}} a_{i}(t)\,b_{i}(t),
\end{align}
where $\{\cdot\}_{s}$ denotes symmetric (Weyl) ordering, $a_{i}(t)$ and $b_{i}(t)$ are the classical counterparts of $\hat{A}$ and $\hat{B}$ in the $i$th trajectory, and $n_{t}$ is the number of trajectories.
\begin{figure}
 \includegraphics[width=1\columnwidth]{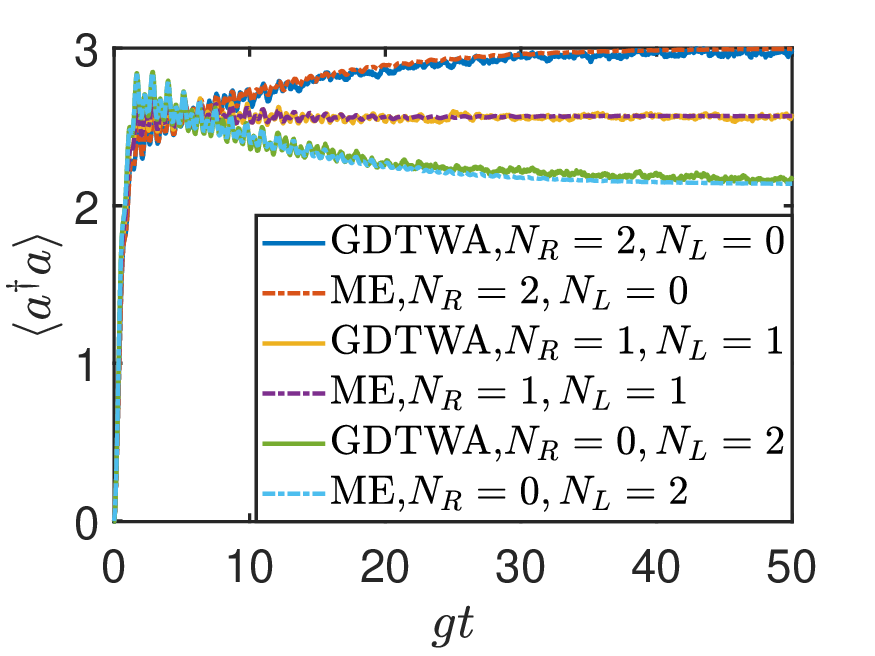}
 \caption{Comparison between the exact numerical results of the master equation and the GDTWA under simulations with 10000 trajectories. The parameters are set as $\Delta_{31}=\Delta_{32}=\Delta_{c}=0,\Omega_{32}=5g,\Omega_{31}=g,\kappa=5g,\eta=4g,\phi_{L}=0$, and $\phi_{R}=\pi$.}
 \label{contrast}
\end{figure}

To benchmark the applicability of GDTWA to our model, we compare its prediction for the intracavity photon number dynamics with exact master equation results for $N_L+N_R=2$; see Fig.~\ref{contrast}. The initial state places the cavity in vacuum and prepares all molecules in $\ket{3^{Q}_{n_{Q}}}$. The GDTWA data averaged over $10^{4}$ stochastic trajectories closely track the exact dynamics over the full time window, thereby validating the method for larger systems with $N_L+N_R\gg 2$. In addition, for different compositions $(N_L,N_R)$ we observe distinct steady-state photon numbers, motivating an enantiodetection protocol based solely on steady-state photon-number readout.
\subsubsection{Interference Mechanism}
Before discussing enantiodetection, we first analyze the case in which the cavity couples to multiple molecules of identical chirality, taking left-handed molecules as an example.

\begin{figure}
 \includegraphics[width=1\columnwidth]{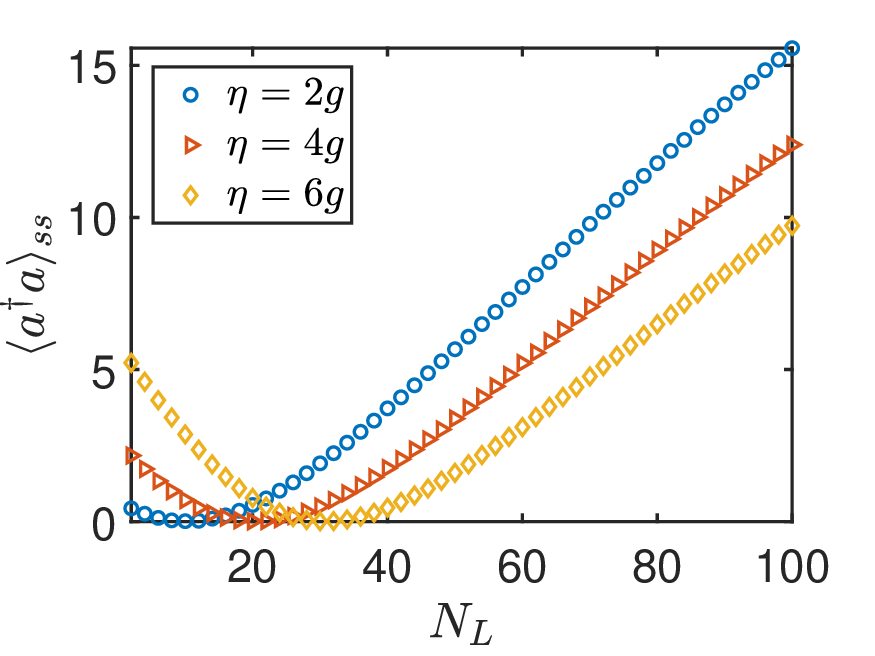}
 \caption{Steady-state photon number as a function of the number of left-handed molecules $N_{L}$ for different strengths $\eta$. The other parameters are set as $\Delta_{31}=\Delta_{32}=\Delta_{c}=0,\Omega_{32}=5g,\Omega_{31}=g,\kappa=5g,\phi_{L}=0$, and $N_{R}=0$.}
 \label{Left}
\end{figure}
Fig.~\ref{Left} shows the steady-state intracavity photon number $\langle\hat{a}^{\dagger}\hat{a}\rangle_{\mathrm{ss}}$ as a function of the molecule number $N_{L}$. The curves exhibit a nonmonotonic dependence at drive strength, and the location of the zero point $\langle\hat{a}^{\dagger}\hat{a}\rangle_{\mathrm{ss}}=0$ shifts to larger molecule numbers as the drive amplitude $\eta$ increases.

In a path-interference picture, a zero of $\langle\hat{a}^{\dagger}\hat{a}\rangle_{\mathrm{ss}}$ signal completely comes from the destructive interference between two molecule-mediated pathways (after eliminating the molecular degrees of freedom):
\begin{align*}
&\text{1. direct drive:}\quad \ket{n-1}\xrightarrow{\ \eta\ } -i\,\ket{n},\\
&\text{2. molecule-mediated:}\quad \ket{n-1}\xrightarrow{\ \Omega_{32},\,g,\,\Omega_{31}\ } i\,\ket{n}.
\end{align*}
As the drive is increased, the amplitude of the direct pathway grows; the observed rightward shift of the zero thus implies that the effective molecule-mediated-pathway amplitude increases with the number of molecules, so a larger $N_{L}$ is required to balance the stronger drive. This collective, interference-based mechanism underlies enantiodetection in a mixture.

We now consider mixture containing both two enantiomers in the presence of the coherent drive. In this case, three effective photonic pathways contribute:
\begin{align}
&1.\ \text{drive:}\qquad \ket{n-1}\xrightarrow{\ \eta\ } -i\,\ket{n},\nonumber\\
&2.\ \text{left-handed:}\quad \ket{n-1}\xrightarrow{\ \Omega_{32},\,\Omega_{31},\,g\ } i\,\ket{n},\label{mix_path}\\
&3.\ \text{right-handed:}\ \ \ket{n-1}\xrightarrow{\ \Omega_{32},\,\Omega_{31},\,g\ } -i\,\ket{n}.\nonumber
\end{align}
It is clear that the left-handed (right-handed) molecular-mediated pathway interferes destructively (constructively) with the drive-induced pathway, while the two molecular-mediated pathways mutually interfere destructively. Meanwhile, since the amplitudes of the molecule-mediated pathways scale with the respective species populations, the steady-state photon number depends on both the drive strength and the enantiomeric excess $\mathcal{P}$.

\begin{figure}
 \includegraphics[width=1\columnwidth]{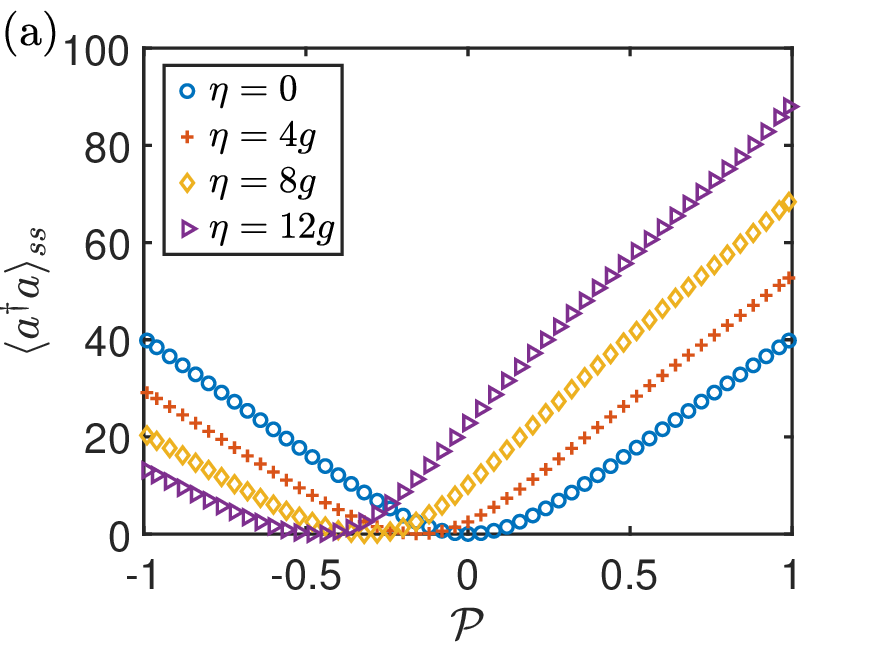}\\
 \includegraphics[width=1\columnwidth]{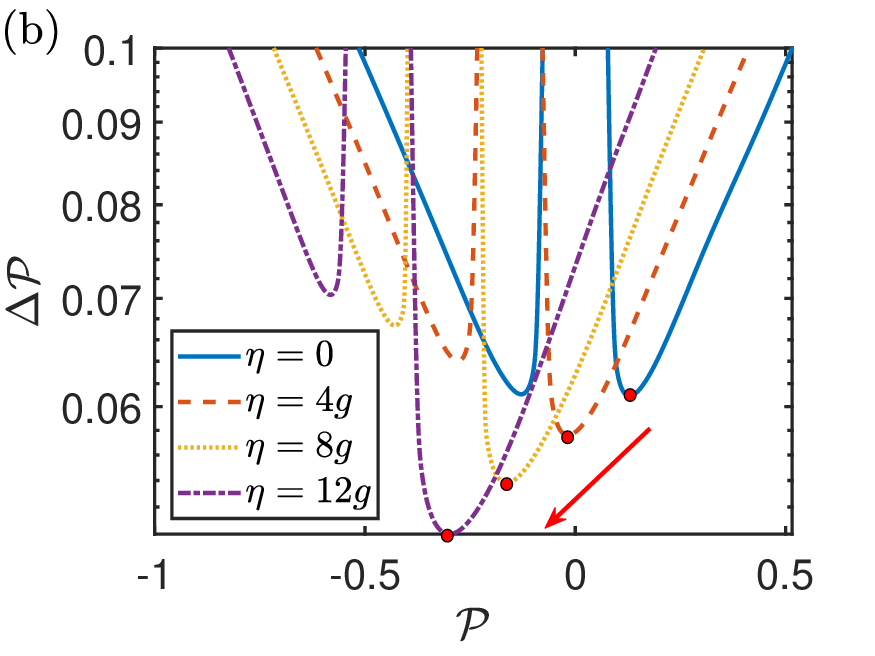}
 \caption{(a) and (b) Steady-state intracavity photon number $\langle \hat{a}^{\dagger}\hat{a}\rangle_{\mathrm{ss}}$ and detection uncertainty $\Delta\mathcal{P}$ as functions of enantiomeric excess $\mathcal{P}$ and several drive strengths $\eta$. The red circles mark the minimum uncertainty for each $\eta$ in panel (b). Other parameters we set as $\Delta_{31}=\Delta_{32}=\Delta_{c}=0, \Omega_{32}=5g, \Omega_{31}=g, \kappa=5g, \phi_{L}=0,\phi_{R}=\pi$, and $N_{R}+N_{L}=200$.}
 \label{Measurement}
\end{figure}
Fig.~\ref{Measurement}(a) shows the steady-state photon number as a function of $\mathcal{P}$ for a fixed total molecule number $N_{L}+N_{R}=200$. With no external drive ($\eta=0$), only the two molecule-mediated pathways interfere, complete cancellation occurs only at equal populations, hence $\langle\hat{a}^{\dagger}\hat{a}\rangle_{\mathrm{ss}}=0$ at $\mathcal{P}=0$. In the presence of a drive ($\eta\neq 0$), the zero shifts to $\mathcal{P}<0$, reflecting that a larger left-handed fraction is required to destructively balance the drive pathway. As the drive strength increases, this zero moves to more negative $\mathcal{P}$.

On the right side of the zero, the slope $|\partial\langle\hat{a}^{\dagger}\hat{a}\rangle_{\mathrm{ss}}/\partial\mathcal{P}|$ is relatively large, enabling lower-error estimation of $\mathcal{P}$ in that region. Using standard error propagation,
\begin{align}
\Delta\mathcal{P}
=\frac{\sqrt{\langle(\hat{a}^{\dagger}\hat{a})^{2}\rangle_{\mathrm{ss}}
-\langle\hat{a}^{\dagger}\hat{a}\rangle_{\mathrm{ss}}^{2}}}
{\bigl|\partial\langle\hat{a}^{\dagger}\hat{a}\rangle_{\mathrm{ss}}/\partial\mathcal{P}\bigr|},
\end{align}
we obtain the detection uncertainty $\Delta\mathcal{P}$ versus $\mathcal{P}$, shown in Fig.~\ref{Measurement}(b). As $\eta$ increases, the point of minimal error shifts toward smaller (more negative) $\mathcal{P}$ and the minimal error decreases. In practice, once a rough range of $\mathcal{P}$ is known, one can choose $\eta$ to place operation near the steep-slope region for high-precision readout. For strong driving, e.g., $\eta=12g$, the minimal error can fall below $5\%$.

When right-handed molecules dominate ($\mathcal{P}>0$) in the mixture, applying a $\pi$ phase shift to the drive (effectively $\eta\to-\eta$) reverses the interference relation between the drive and molecular-mediated pathways, moving the zero of $\langle\hat{a}^{\dagger}\hat{a}\rangle_{\mathrm{ss}}$ to $\mathcal{P}>0$ and shifting the corresponding optimal detection window accordingly.
\section{conclusion}
In conclusion, we have investigated enantiodetection for single molecules and finite-size mixtures in a cavity-QED platform using a quantum-optical approach. By leveraging the chirality-dependent loop phase of cyclic three-level ($\Delta$-type) molecules, we engineer interference between drive-induced and molecule-mediated pathways. This mechanism establishes a clear chirality dependence of the steady-state intracavity photon number, enabling single-molecule enantiodetection via a simple photon-number readout.

For finite ensembles, we showed that the amplitude of the molecule-mediated pathway increases with the number of molecules, producing an enantiomeric-excess dependence of the steady-state photon number that can be tuned by the drive strength. In the optimal operating regime, the resulting protocol achieves errors below $5\%$. Methodologically, we employed the GDTWA to capture finite-size correlations and fluctuations beyond mean field, providing a scalable framework for enantiodetection in mesoscopic molecular mixture and offering a practical route to quantum-optical sensing of chirality.
\section*{acknowledgments}
This work is supported by the Quantum Science and Technology-National Science and Technology Major Project (No. 2023ZD0300700), National Science Foundation of China (Grant Nos. 12375010 and 12574387) and the Hainan Provincial Natural Science Foundation of China (No. 125RC631).

\appendix
\begin{widetext}
\addcontentsline{toc}{section}{Appendices}\markboth{APPENDICES}{}
\section{generalized discrete truncated Wigner approximation}
In this Appendix, we provide details of the generalized discrete truncated Wigner approximation (GDTWA).

For a system with $D$ discrete states $\{\ket{1},\ket{2},\dots,\ket{D}\}$, the density operator can be expanded in the identity operator $I$ and the set of generalized Gell-Mann matrices (GGMs). There are $(D^{2}-1)$ GGMs, defined as
\begin{align}
&\bm{\Lambda}_{\alpha,\beta<\alpha}^{R}
=\ket{\beta}\bra{\alpha}+\ket{\alpha}\bra{\beta}
\qquad (1\le \alpha\le D,\ 1\le \beta\le \alpha-1),\nonumber\\
&\bm{\Lambda}_{\alpha,\beta<\alpha}^{I}
=-i\big(\ket{\beta}\bra{\alpha}-\ket{\alpha}\bra{\beta}\big)
\qquad (1\le \alpha\le D,\ 1\le \beta\le \alpha-1),\nonumber\\
&\bm{\Lambda}_{\alpha}^{D}
=\sqrt{\frac{2}{\alpha(\alpha+1)}}
\!\left(\sum_{\beta=1}^{\alpha}\ket{\beta}\bra{\beta}
-\alpha\,\ket{\alpha+1}\bra{\alpha+1}\right)
\qquad (1\le \alpha\le D-1),
\end{align}
which are Hermitian operators (system observables). They are traceless, $\mathrm{Tr}(\bm{\Lambda}_{\mu})=0$, and obey the orthogonality relation $\mathrm{Tr}(\bm{\Lambda}_{\mu}\bm{\Lambda}_{\nu})=2\delta_{\mu\nu}$. Under these conventions, the density operator of a $D$-level system can be written as
\begin{equation}
\hat{\rho}
=\frac{1}{D}\left(I+\sum_{\mu}\frac{D}{2}\,\langle \bm{\Lambda}_{\mu}\rangle(t)\,\bm{\Lambda}_{\mu}\right),
\end{equation}
so that knowledge of the expectation values $\langle \bm{\Lambda}_{\mu}\rangle(t)$ uniquely specifies $\hat{\rho}$.

In our model, the master equation contains coherent evolution generated by the Hamiltonian in Eq.~(\ref{H}) and an incoherent part due to cavity loss,
\begin{align}
\frac{d\hat{\rho}}{dt}
=-i[\hat{H},\hat{\rho}]
+\frac{\kappa}{2}\big(2\hat{a}\hat{\rho}\hat{a}^{\dagger}
-\hat{a}^{\dagger}\hat{a}\hat{\rho}
-\hat{\rho}\hat{a}^{\dagger}\hat{a}\big).
\end{align}
Within a mean-field closure, the dynamics of the GGM averages for $n_{Q}$th molecule (with $Q=L,R$ and $n_{Q}=1,2,\dots,N_{Q}$) obey the following set of equations:
\begin{align}
&\frac{d\lambda_{2,1}^{(n_{Q},R)}}{dt}=(\Delta_{31}-\Delta_{32})\lambda_{2,1}^{(n_{Q},I)}+2g\lambda_{1}^{(n_{Q},D)}\text{Im}[\alpha]+\Omega_{32}\left(\lambda_{3,1}^{(n_{Q},I)}\cos(\phi_{Q})-\lambda_{3,1}^{(n_{Q},R)}\sin(\phi_{Q})\right)+\Omega_{31}\lambda_{3,2}^{(n_{Q},I)},\nonumber\\
&\frac{d\lambda_{2,1}^{(n_{Q},I)}}{dt}=-(\Delta_{31}-\Delta_{32})\lambda_{2,1}^{(n_{Q},R)}-2g\lambda_{1}^{(n_{Q},D)}\text{Re}[\alpha]-\Omega_{32}\left(\lambda_{3,1}^{(n_{Q},R)}\cos(\phi_{Q})+\lambda_{3,1}^{(n_{Q},I)}\sin(\phi_{Q})\right)+\Omega_{31}\lambda_{3,2}^{(n_{Q},R)},\nonumber\\
&\frac{d\lambda_{3,1}^{(n_{Q},R)}}{dt}=\Delta_{31}\lambda_{3,1}^{(n_{Q},I)}-g\left(\lambda_{3,2}^{(n_{Q},R)}\text{Im}[\alpha]+\lambda_{3,2}^{(n_{Q},I)}\text{Re}[\alpha]\right)+\Omega_{32}\left(\lambda_{2,1}^{(n_{Q},R)}\sin(\phi_{Q})+\lambda_{2,1}^{(n_{Q},I)}\cos(\phi_{Q})\right),\nonumber\\
&\frac{d\lambda_{3,1}^{(n_{Q},I)}}{dt}=-\Delta_{31}\lambda_{3,1}^{(n_{Q},R)}+g\left(\lambda_{3,2}^{(n_{Q},R)}\text{Re}[\alpha]-\lambda_{3,2}^{(n_{Q},I)}\text{Im}[\alpha]\right)-\Omega_{32}\left(\lambda_{2,1}^{(n_{Q},R)}\cos(\phi_{Q})-\lambda_{2,1}^{(n_{Q},I)}\sin(\phi_{Q})\right)\nonumber\\
&-\Omega_{31}(\lambda_{1}^{(n_{Q},D)}+\sqrt{3}\lambda_{2}^{(n_{Q},D)}),\nonumber\\
&\frac{d\lambda_{3,2}^{(n_{Q},R)}}{dt}=\Delta_{32}\lambda_{3,2}^{(n_{Q},I)}+g\left(\lambda_{3,1}^{(n_{Q},R)}\text{Im}[\alpha]-\lambda_{3,1}^{(n_{Q},I)}\text{Re}[\alpha]\right)+\Omega_{32}\left(\sqrt{3}\lambda_{2}^{(n_{Q},D)}-\lambda_{1}^{(n_{Q},D)}\right)\sin(\phi_{Q})-\Omega_{31}\lambda_{2,1}^{(n_{Q},I)},\nonumber\\
&\frac{d\lambda_{3,2}^{(n_{Q},I)}}{dt}=-\Delta_{32}\lambda_{3,2}^{(n_{Q},R)}+g\left(\lambda_{3,1}^{(n_{Q},R)}\text{Re}[\alpha]+\lambda_{3,1}^{(n_{Q},I)}\text{Im}[\alpha]\right)-\Omega_{32}\left(\sqrt{3}\lambda_{2}^{(n_{Q},D)}-\lambda_{1}^{(n_{Q},D)}\right)\sin(\phi_{Q})-\Omega_{31}\lambda_{2,1}^{(n_{Q},R)},\nonumber\\
&\frac{d\lambda_{1}^{(n_{Q},D)}}{dt}=-2g\left(\lambda_{2,1}^{(n_{Q},R)}\text{Im}[\alpha]-\lambda_{2,1}^{(n_{Q},I)}\text{Re}[\alpha]\right)+\Omega_{32}\left(\lambda_{3,2}^{(n_{Q},R)}\sin(\phi_{Q})-\lambda_{3,2}^{(n_{Q},I)}\cos(\phi_{Q})\right)+\Omega_{31}\lambda_{3,1}^{(n_{Q},I)},\nonumber\\
&\frac{d\lambda_{2}^{(n_{Q},D)}}{dt}=\sqrt{3}\Omega_{32}\left(\lambda_{3,2}^{(n_{Q},I)}\cos(\phi_{Q})-\lambda_{3,2}^{(n_{Q},R)}\sin(\phi_{Q})\right)+\sqrt{3}\Omega_{31}\lambda_{3,1}^{(n_{Q},I)},\nonumber\\
&\frac{d\alpha}{dt}=-(i\Delta_{c}+\frac{\kappa}{2})\alpha-i\eta-i\frac{g}{2}\sum_{Q=L,R}\sum_{n_{Q}=1}^{N_{Q}}(\lambda_{2,1}^{(n_{Q},R)}+i\lambda_{2,1}^{(n_{Q},I)}).
\label{Eq}
\end{align}
Here, $\lambda_{\mu}=\langle \bm{\Lambda}_{\mu}\rangle$ and $\alpha=\langle \hat{a}\rangle$. A direct integration of Eq.~(\ref{Eq}) with initial conditions $\lambda_{\mu}(0)=\operatorname{Tr}[\bm{\Lambda}_{\mu}\hat{\rho}(0)]$ and $\alpha(0)=\operatorname{Tr}[\hat{a}\hat{\rho}(0)]$ often yields inaccurate dynamics away from the thermodynamic limit or in strongly correlated regimes. The reason is that this procedure evolves a single deterministic trajectory and therefore ignores the quantum uncertainty of the initial state. In finite systems and strongly interacting settings, these initial-state fluctuations substantially influence the evolution, leading to deviations from mean-field predictions. For open systems, additional noise originating from dissipation further perturbs the dynamics and is not captured by a purely deterministic mean-field description.

In GDTWA, the initial values of operator averages are sampled from a Wigner-type quasi-distribution, generating an ensemble of trajectories that encode the correct quantum fluctuations. Concretely, we diagonalize each GGM as
\begin{align}
\bm{\Lambda}_{\mu}=\sum_{a_{\mu}=1}^{D}\lambda_{a_{\mu}}\ket{\eta_{a_{\mu}}}\bra{\eta_{a_{\mu}}},
\end{align}
where $\ket{\eta_{a_{\mu}}}$ is an eigenvector of $\bm{\Lambda}_{\mu}$ with eigenvalue $\lambda_{a_{\mu}}$. The sampling rule sets the probability for the initial value $\lambda_{\mu}(0)=\lambda_{a_{\mu}}$ to
$P_{a_{\mu}}=\operatorname{Tr}[\ket{\eta_{a_{\mu}}}\bra{\eta_{a_{\mu}}}\hat{\rho}(0)]$.
For the cavity field, $\alpha(0)$ is drawn from the Wigner function of the chosen initial optical state; for a coherent state with mean photon number $N$, this corresponds to a Gaussian distribution centered at $\sqrt{N}$ (with the appropriate initial phase).

To model vacuum input noise associated with cavity dissipation, we augment the cavity equation with an additive complex Wiener process,
\begin{align}
d\alpha=\Big[-(i\Delta_{c}+\tfrac{\kappa}{2})\alpha-i\eta-\tfrac{i g}{2}\sum_{Q=L,R}\sum_{n_{Q}=1}^{N_{Q}}\big(\lambda_{2,1}^{(n_{Q},R)}+i\lambda_{2,1}^{(n_{Q},I)}\big)\Big]dt+\frac{\sqrt{\kappa}}{2}\big[dw_{1}(t)+i\,dw_{2}(t)\big],
\end{align}
where $dw_{1}(t)$ and $dw_{2}(t)$ are real, independent Wiener increments over $[t,t+dt]$ with $\langle dw_{i}(t)\rangle=0$ and $\langle dw_{i}(t)dw_{j}(t')\rangle=\delta_{ij}\delta(t-t')\,dt$.

Because the initial sampling uses the Wigner representation, symmetrically ordered operator averages are obtained by trajectory averages of the corresponding classical variables. For arbitrary operators $\hat{A}$ and $\hat{B}$,
\begin{align}
\langle \hat{A}\rangle(t)=\frac{1}{n_{t}}\sum_{i=1}^{n_{t}} a_{i}(t),\qquad
\langle \{\hat{A}\hat{B}\}_{s}\rangle(t)=\frac{1}{n_{t}}\sum_{i=1}^{n_{t}} a_{i}(t)\,b_{i}(t),
\end{align}
where $a_{i}(t)$ and $b_{i}(t)$ are the classical counterparts of $\hat{A}$ and $\hat{B}$ along the $i$th stochastic trajectory, and $n_{t}\gg 1$ ensures adequate sampling of fluctuations and noise.
\end{widetext}

\end{document}